\documentclass[a4paper,12pt,pic98proei]{article}
\usepackage{epsfig}
\usepackage{english}
\begin{document}
\title{ 
Nucleon Structure Functions
}
\author{
F. ~Eisele                           \\
{\em Physikalisches Institut der Universit\"at Heidelberg} 
}
\maketitle
\baselineskip=14.5pt
\begin{abstract}
New structure function measurements from fixed target experiments and especially HERA
 are reviewed. The extraction of parton distributions from these measurements is
 discussed with special emphasis on systematic problems. New information from
Drell-Yan 
 and direct photon production experiments are also presented . Finally the
present uncertainties of our knowledge on parton distributions and on $\alpha _s$
from DIS experiments are discussed \footnote{Invited talk XVIII Physics in Collision}.
\end{abstract}
\baselineskip=17pt

%
\section{Introduction}
\begin{figure}[htbt] \begin{center}
\epsfig{file=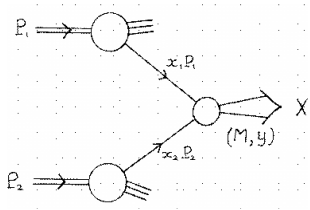,width=4.cm,clip=}
 \caption{\it
     Example of a LO hard scattering process in hadron-hadron scattering.
    \label{hardprocess} }
\end{center} \end{figure}

Our knowledge of nucleon structure functions and of the parton distributions (PDF's)
derived
 from them has steadily improved due to both the improvement of a large variety of
 measurements and a more sophisticated theoretical treatment of hard scattering 
 processes
  in perturbative QCD. Structure functions and PDF's are needed for two reasons:
\begin{itemize}
\item They are a necessary input for all hard scattering processes involving
 nucleons
to make precise predictions in the standard model and of course to look for 
deviations
 from theses predictions, and to make predictions for signals and backgrounds at
 colliders especially the LHC.
\item They contain important information about the underlying physics of hadrons 
and 
allow stringent tests of perturbative QCD.
\end{itemize}
A hard scattering process in a hadron hadron collision is shown in figure
 \ref{hardprocess}.
 Two partons with momentum fractions $x_1$ and $x_2$ scatter
 and produce a heavy state with mass M which can be jets, $\gamma$+jet, heavy Bosons
 or Quarks or new particles like Higgs or SUSY. The differential
 parton-parton cross
section $d\sigma_{ij}$  is given by pQCD, the total cross section is given by 
$$d\sigma = \sum_{i,j}\int dx_1 dx_2 f_i(x_1,\mu ^2) f_j  (x_2,\mu ^2) \star
 d\sigma_{ij} (p_1,p_2, \alpha_s(\mu ^2), M_2/\mu ^2) $$
where the sum goes over all parton flavours which contribute.
 In LO pQCD the cross section factorises into
 the parton luminosity which depends on the parton densitites
$f_i, f_j$ and the differential parton-parton cross section. The parton densities 
are therefore universal to all hard scattering processes provided they are extracted
in higher order (so far in NLO) and corrected to the leading order diagrams. The
parton densities depend on the fractional momentum x and the typical scale $\mu^2$
of the process.
\par
 Deep inelastic lepton nucleon scattering (DIS) is the basis for 
our
 knowlegde
of PDF's.  The kinematic plane of available measurements in (x,$Q^2$) from
 fixed
\begin{figure}[htbt] \begin{center}
\epsfig{clip=,width=8.cm,file=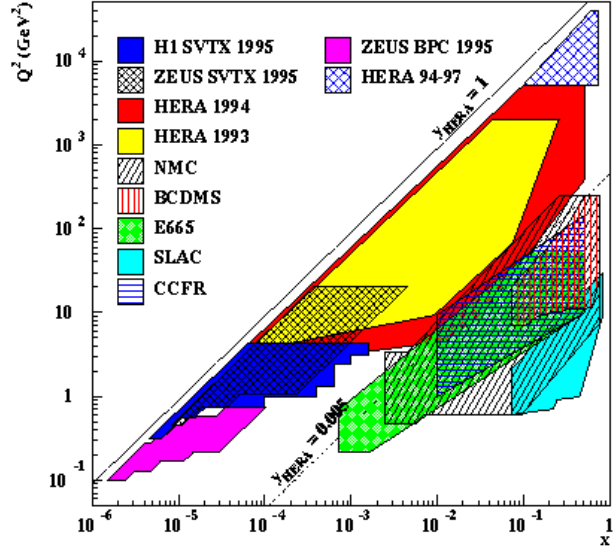}
 \caption{\it
  Available DIS data from fixed target and HERA.
    \label{DISplane} }
\end{center} \end{figure}
 target and HERA experiments is shown in figure \ref{DISplane}.

\begin{figure}[htbt] \begin{center}
\epsfig{clip=,width=7.5cm,file=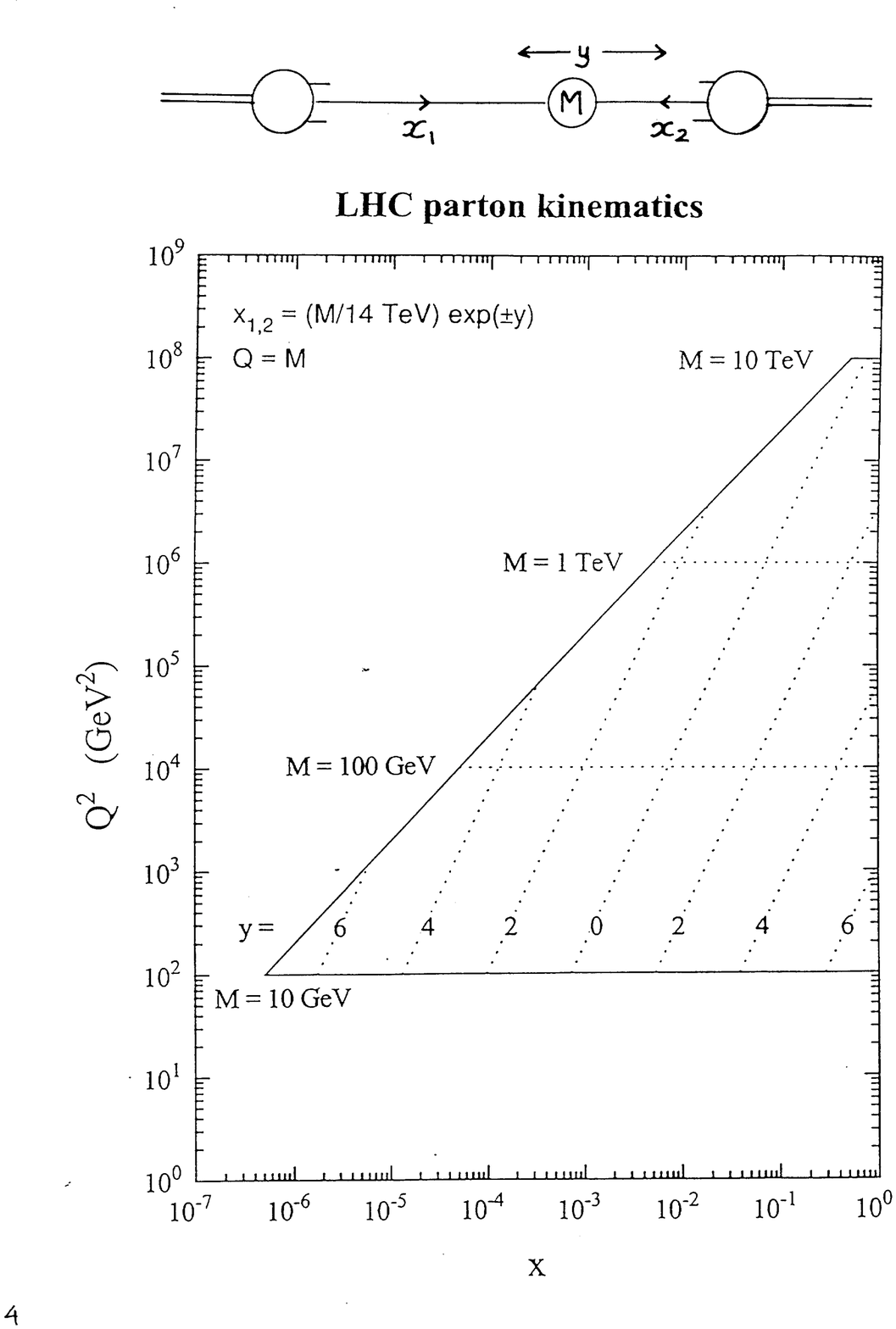 }
 \caption{\it
 Kinematic plane of hard  scattering processes at LHC. Lines of constant rapidity y 
for the produced system with mass M are given as dashed lines.
    \label{LHCplane} }
\end{center} \end{figure}

 The data
 cover now a huge region  
$$ 2 \star 10^{-5} < x < .75 
 \ ; \  1 \  GeV^2 < Q^2 < 40000 \ GeV^2 $$
where HERA has added 2 orders of magnitude in both variables. The kinematic
plane for hard scattering processes at LHC (14 TeV) is compared in figure
\ref{LHCplane} where the mass M of the produced system is used as scale.
 Also
shown are lines of constant rapidity y of the system M which is proportional
 to ($x_1 - x_2$. It is obvious that due to the high CMS energy of LHC a very
 large fraction
of interesting LHC physics is physics at low x ( here taken as
$x<10^{-2}$) \cite{stirling}. For example the production of Higgs particles in the mass range
$ 100 \ GeV < M_H < 500 \ GeV $ in the rapidity range $ |y| < 2.5 $ requires the
knowledge of the gluon density for x-values as low as $10^{-4}$ which is  now
available due to HERA. \par
In order to apply our present knowledge on PDF's we have however to
extrapolate the parton distribution by up to 3 orders in magnitude in $Q^2$.
This is possible using the pQCD evolution equations. At large x the DGLAP
equations are expected to provide a sufficiently good approximation. This is
however rather doubtful at small x because terms proportional to $\alpha_s
\star ln1/x $ have been neglected in the derivation of DGLAP compared to the 
$\alpha_s \star lnQ^2 $ terms which have been summed. This does not look
justified at small x. One important question at HERA is therefore the test of
QCD dynamics at small x.
\clearpage
\section{New DIS data}
\begin{figure}[htbt] \begin{center}
\epsfig{clip=,width=8.cm,file=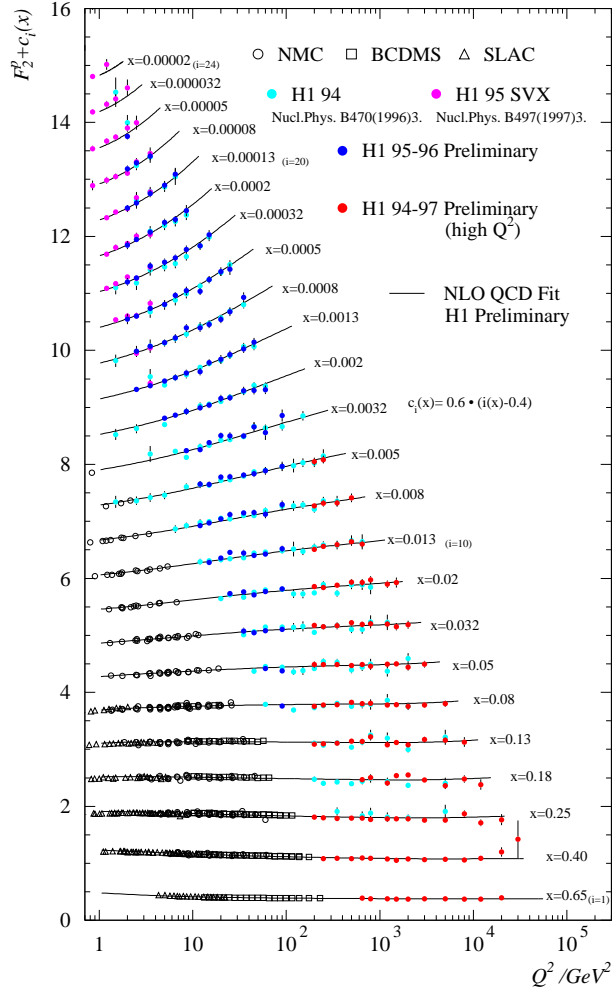}
 \caption{\it
  The structure function $F_2^{lp} \ vs. \ Q^2 $ for fixed values of x for recent HERA 
 and for fixed target data. The solid lines are the result of a NLO pQCD fit to these
 data by H1 as discussed below.
    \label{F2epall} }
\end{center} \end{figure}
Progress in the measurement of nucleon structure functions is steady but
rather slow. It takes long time and hard work to make reliable measurements
and to study and reduce sytematic effects. Nevertheless the best measurements
are all dominated by correlated systematic errors which makes it difficult to 
analyse these data in a consistent quantitative way. From fixed target
experiments two important new results have been added. The NMC collaboration
has
published the final analysis of their data taken back in '89 with a small
angle trigger \cite{NMC}. These data enlarge the low x range and therefore the overlap
with HERA  and also allow to measure $R= \sigma _L / \sigma _T $.
 The CCFR 
neutrino collaboration at Fermilab has published an improved analysis of their
data first published in '93 \cite{CCFR}. They have essentially improved the 
energy calibration for hadrons and muons. Their structure function
measurements have a significant impact on our knowledge of $\alpha_s$, to be
discussed at the end. 
\begin{figure}[htbt] \begin{center}
\epsfig{clip=,width=5.cm,file=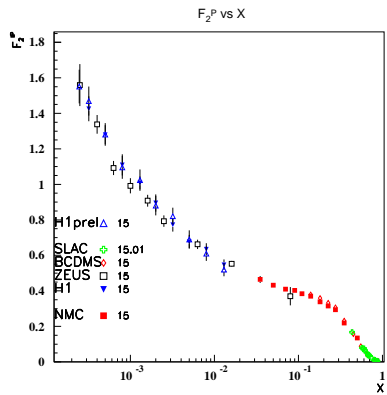}
 \caption{\it
  Fixed target and HERA measurements of  $F_2^{lp}$ for a fixed value of
  $Q^2=15 \ GeV^2$.
    \label{f215GeV} }
\end{center} \end{figure}
\begin{figure}[htbt] \begin{center}
\epsfig{clip=,width=5.cm,file=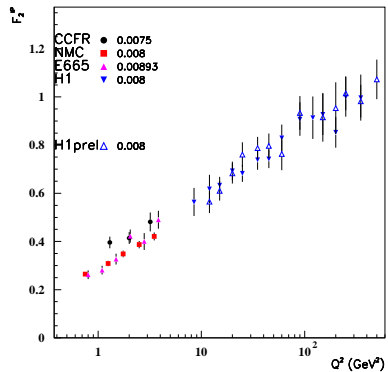 }
 \caption{\it
 Fixed target and HERA measurements of  $F_2^{lp}$ for a fixed value of x=.008.
    \label{f2x.008} }
\end{center} \end{figure}
The H1 collaboration has made a very substantial
step towards large x and $Q^2$  which allows for the first time to directly
test the $Q^2$-evolution at large x over two orders of magnitude with
reasonable precision. Also more precise data in the low x
region has been provided by both H1 and ZEUS.
A compilation of some of the most precise measurements of
the structure function $F_2^{ep}$ is shown in figure \ref{F2epall} versus
$Q^2$ for fixed values of x.
 Figures \ref{f215GeV} and
\ref{f2x.008} show all available data points for a fixed value of $Q^2 = 15 \ GeV^2$
 resp. a fixed value of x=.008 in order to see the relative precision
and the matching of fixed target and HERA data. 
 Figure \ref{f215GeV}
demonstrates very clearly one of the most important  contributions of HERA so far: The
structure function at small x and therefore the $q \bar q $
 sea rises by more
than a factor 3 in the HERA range towards small x. This has severely affected
the predictions for processes at LHC both for pp and heavy ion collisions. The
structure function for fixed x=.008 rises by a factor 5 with $Q^2$
demonstrating huge scaling violations whch have to be explained by pQCD. The
overall agreement among the two HERA experiments and the matching with fixed target
experiments is good. Our knowledge on $F_2^{ep}$
 has presently a
systematic uncertainty  of about $\pm 3 \%$ over the whole x-range $10^{-4} <x < .65$
 mainly
limited by the uncertainties of energy scales.
%
\section{ Analysis of the low x region and the gluon distribution}
The low x data on $F_2$ 
show large scaling violations .
\begin{figure}[htbt] \begin{center}
\epsfig{clip=,width=7.cm,file=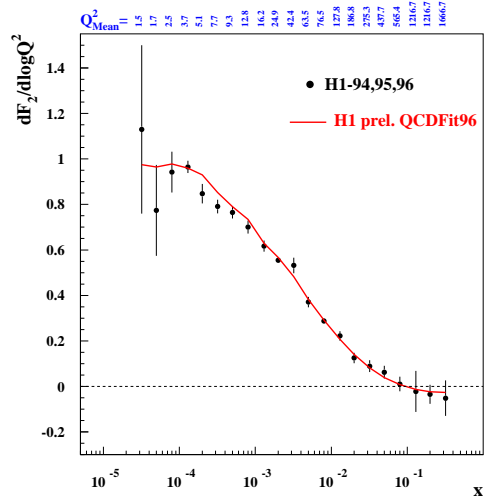 }
 \caption{\it
Measured slopes $dF_2  / dlnQ^2$ obtained from linear fits in ln$Q^2$ to the H1 data.
The solid line gives the slopes as obtained from the NLO QCD fit.
    \label{slopeslowx} }
\end{center} \end{figure}
 For large enough $Q^2$
these should be described by the pQCD evolution equations:
$$ d\Sigma (x,Q^2)/dlnQ^2 = \alpha_s /2\pi \ (P_{gq} \otimes g \ + \ P_{qg} \otimes
  \ 
\Sigma ) $$
$$ dxg(x,Q^2)/dlnQ^2 = \alpha_s /2\pi \ (P_{gg} \otimes g \ + \ P_{qg} \otimes
 \Sigma
) $$
Here $\Sigma =\sum_f x(q_f + \bar q_f) $ is the sum of all quark and antiquark
momentum distributions.  The parton distributions evolve with $Q^2$ because
quarks radiate gluons, gluons split into quar-antiquark pairs and gluons split
into gluons. At small x the gluon radiation from quarks can be neglected. The
evolution is then dominated by the evolution of the gluon distribution.
 The splitting
functions $P_{ij}$ depend on terms proportional to
$$ \alpha_s \star lnQ^2 , \alpha_s \star lnQ^2 ln1/x \ and \ \alpha_s \star ln1/x $$
 If terms proportional ln1/x  are neglected, then we arrive at the DGLAP
evolution equation, if instead terms proportional to ln1/x are kept instead of
ln$Q^2$ terms we arrive at the BFKL equation.

 The first thing to do is therefore
to check the evolution directly. Figure \ref{slopeslowx} compares the measured
slopes $dF_2 /dlnQ^2 \ (x)$ (by linear fits in ln$Q^2$ for fixed x ) with the
predictions of a DGLAP fit to the data ( H1 fit discussed below). There is good
agreement down to $x=10^{-4}$. This illustrates, to our big surprise, that the DGLAP
aproximations works even at very small values of x. Other attempts to see 'BFKL' like
 effects in the final state , which are not discussed here, were also negative.\par
This observation  gives us the possibility to measure the gluon distribution from the
measured slopes  $dF_2 /dlnQ^2 \ (x)$. To rather good approximation the relation at
 small x is given by:
$$dF_2 /dlnQ^2 \ (x) \approx 10/27 \pi \star \alpha_s (Q^2) \star xg(2x,Q^2)$$
The HERA experiments therefore measure the gluon distribution at small x which is of
 prime importance for physics at LHC.\par%
\begin{figure}[htbt] \begin{center}
\epsfig{clip=,width=8.cm,file=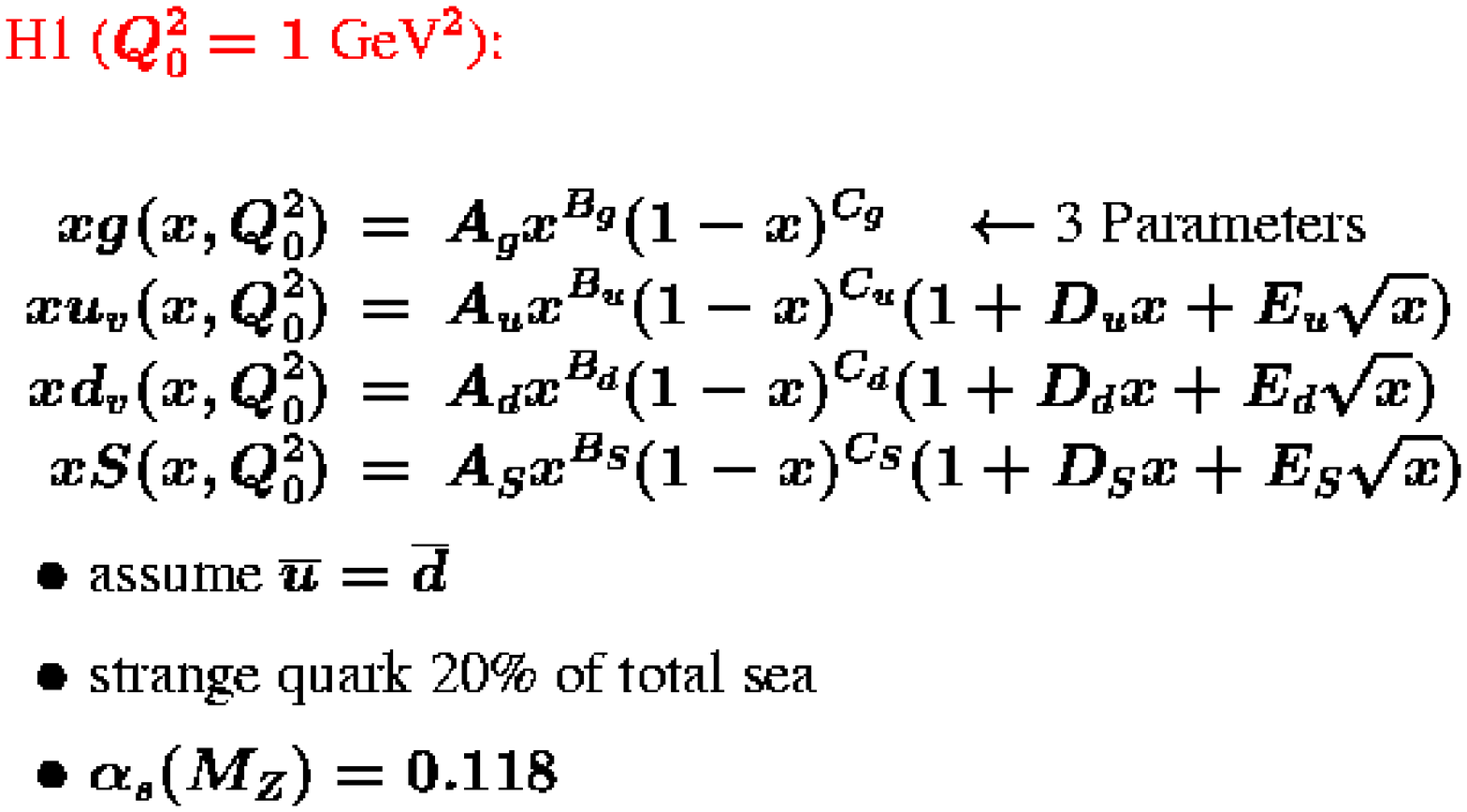 }
 \caption{\it
Parametrisations of parton momentum distributions as used by H1 and choice of parameters.
The solid line gives the slope as obtained from the NLO QCD fit.
    \label{qcdfit} }
\end{center} \end{figure}
\begin{figure}[htbt] \begin{center}
\epsfig{clip=,width=6.cm,file=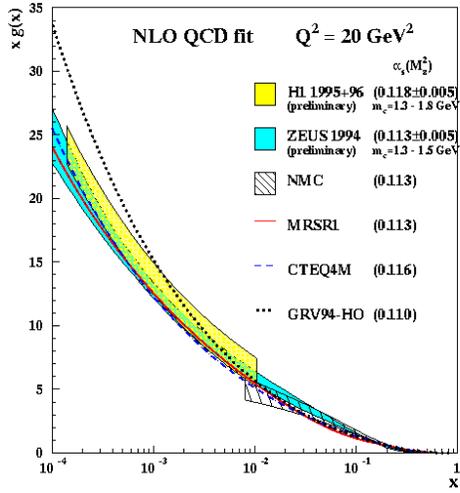}
 \caption{\it
Resulting gluon distribution for $Q^2_0 = 20 GeV^2$ from H1, ZEUS and NMC NLO pQCD fits.
The error bands include the estimate of correlated systematic errors.
    \label{gluonfit} }
\end{center} \end{figure}
\begin{figure}[htbt] \begin{center}
\epsfig{clip=,width=5.cm,file=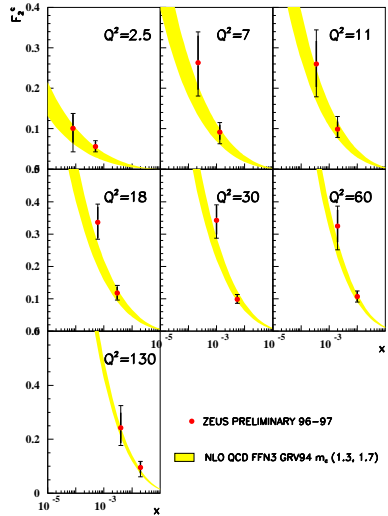 }
 \caption{\it
$F_2^{charm}$ as measured by the ZEUS collaboration using $D^*$ production in DIS compared
 to the NLO prediction based on the $\gamma - gluon$ fusion process.
    \label{f2charm} }
\end{center} \end{figure}
Both HERA collaborations have  determined the gluon distribution from a NLO
DGLAP fit to their data ($Q^2 > 1.5 \ GeV^2$) in combination with ep and ed data from
 BCDMS and NMC. This is explained below for the example of H1 \cite{gluonh1zeus}
which  uses 15 free parameters plus some relative normalisation parameters.
The resulting gluon distribution is shown in figure \ref{gluonfit} at $Q^2 =
20 \ GeV^2$
together with the corresponding determination from ZEUS \cite{gluonh1zeus}. 
The gluon density rises strongly
towards small x and xg(x) reaches values of 25 at $x = 10^{-4}$, a factor 15 larger than
the sea quark density. This underlines the predominace of the gluon evolution by
gluon splitting whereas the splitting of a gluon into a $q\bar q$ pair is relatively rare.
The error bands given include all known strongly correlated systematic errors. The
gluon distribution in the x- range between $10^{-2} \ and \ 10^{-4}$ is therefore known to
about 15\% from HERA. It should be noted that the charmed quark contribution to $F_2$ 
is very substantial (up to 20\%) at small x. This contribution is calculated
directly in NLO pQCD 
\cite{charmpred} via the photon gluon
 fusion process. This contribution has been directly  measured by both HERA experiments
using  DIS events with $D^*$'s. The preliminary result of ZEUS is shown in figure
\ref{f2charm} together wih the pQCD prediction based on the gluon distribution of figure
\ref{gluonfit}.
 There is good agreement showing the consistency of the procedure. 
\clearpage
\section{Transition to low $Q^2$ at HERA}
\begin{figure}[htbt] \begin{center}
\epsfig{clip=,width=7.cm,file=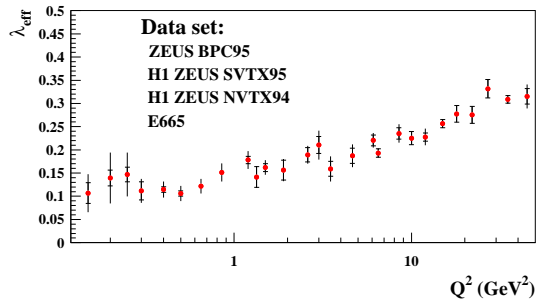 }
 \caption{\it
The exponent $\lambda $ from a fit $F_2 \sim x^{\lambda }$ to the H1 data at small x vs.
$Q^2$.
    \label{lambda} }
\end{center} \end{figure}

\begin{figure}[htbt] \begin{center}
\epsfig{clip=,width=7.5cm,file=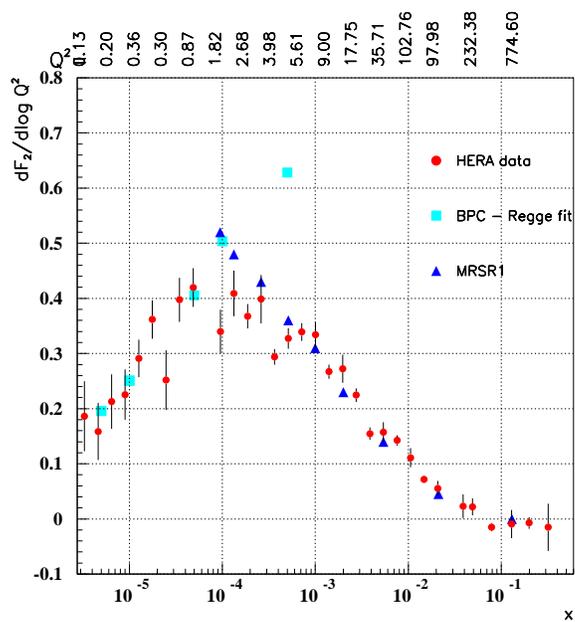}
 \caption{\it
Slopes $dF_2 /dln Q^2$ measured in the transition region compared to predictions of
 a DGLAP fit and of a Regge fit.
    \label{df2dlnq2.lowq2} }
\end{center} \end{figure}

So far we have discussed the 'DIS-regime' where we beleive pQCD is applicable
 without special justification. It's worthwhile to study the transition region from
the DIS regime to the region down to $Q^2 = 0$ , e.g. 'photoproduction'. This has
been possible at HERA due to the installation of special low angle electron
calorimeters which allow to measure inelastic neutral current scattering down to
 $Q^2= 0.11 \ GeV^2$ \cite{ZEUSlowq2}.
\begin{figure}[htbt] \begin{center}
\epsfig{clip=,width=7.cm,file=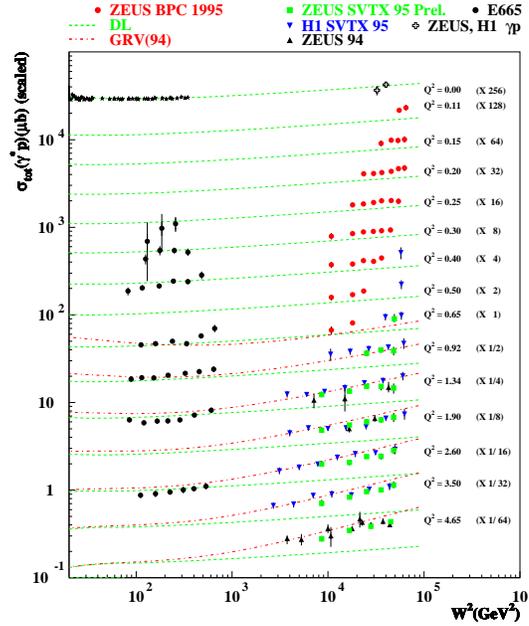 }
 \caption{\it
The total $\gamma ^* p $ cross section vs. $W^2$ for low values of $Q^2$ compared to 
DGLAP (GRV) and soft pomeron regge (DL) predictions.
    \label{sigmagammap} }
\end{center} \end{figure}
 In addition we can of course measure
 the total
 photoproduction
cross section $\sigma^{\gamma p}_T(W^2,Q^2=0)$. The theoretical concepts used to
 describe DIS and photoproduction can be easily related. The structure function
 $F_2(x,Q^2)$ and the $\gamma p $ cross section are related by:
$$ \sigma^{\gamma ^* p}_T(W^2,Q^2) \simeq 4 \pi \alpha^2 /Q^2 F_2^{ep}(x,Q^2) $$
$$ W^2_{\gamma p} \simeq Q^2/x  \  at \ low \ x $$
Scattering of electrons at low x is therefore equivalent to high energy 
$\gamma p$ scattering. Since photons behave like hadrons if they interact with
 nucleons it should not be surprising that the concepts which are used to
describe high energy hadron-hadron scattering are suitable to discuss the
low x region. Figure \ref{sigmagammap} shows $ \sigma^{\gamma ^* p}_T$ as
funtion of $W^2$ for different values of $Q^2$.
 Its well known that the
 photoproduction cross section has the same rise  with $W^2$ as observed in
p-p scattering:
$ \sigma^{\gamma p}_T(W^2,Q^2=0) \sim (W^2)^{\lambda}$ where $\lambda=0.08 $ is
the intercept of the 'soft Pomeron trajectory'. The HERA data at low $Q^2$ show that the
 increase
 of $\sigma_T$ with $W^2$
becomes larger with increasing $Q^2$, therefore $\lambda $ rises with 
 $Q^2$  as shown in figure \ref{lambda} from a value close to .08 at low $Q^2$ to 
$\lambda \approx .35$ at $Q^2=40 GeV^2$.
 The measurements are compared to a soft pomeron 
model of Donnachie and Landshoff (DL, dashed curves) and a pQCD fit based on
the DGLAP evolution (dash-dotted line). We see a smooth transition from regge
 models to the DIS description around $Q^2= .8 \ GeV^2$. Figure \ref{df2dlnq2.lowq2}
shows another interesting aspect of the transition region. The slopes $dF_2/dlnQ^2$
have been directly determined from HERA data 
as a function of x.
 These slopes rise
 towards low x from $x=10^{-1} \ to \ x= \ 10^{-4}$ in agreement with DGLAP
predictions given by the triangles but then fall again towards even lower x.
The slopes at lowest x are in agreement with regge fits based on the 'soft
Pomeron' concept. The transition occurs around $x=10^{-4} \ at \ Q^2 \approx
1 GeV^2$. 
\clearpage
\section{Large x and large $Q^2$}
\begin{figure}[htbt] \begin{center}
\epsfig{clip=,width=7.cm,file=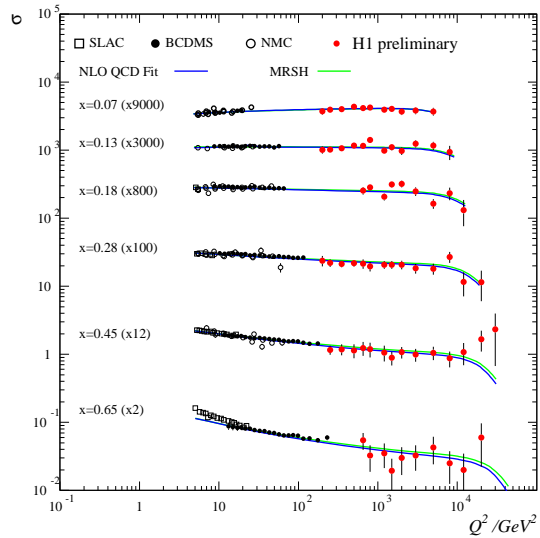 }
 \caption{\it
The reduced cross section for large values of x vs. $Q^2$ compared to two DGLAP fits to
 data with $Q^2 < 120 \ GeV^2$ and inclusing the large x data.
    \label{highxvsq2} }
\end{center} \end{figure}

This is the kinematic region where we want to test the standard model by looking 
for e.g. contact interactions or new massive particle production (see talk by
J. Meyer at this conference). The questions here are;
\begin{itemize}
\item how reliable is the standard model prediction e.g. how well can we 
extrapolate parton densities from fixed target experiments to HERA.
\item what are the experimental limitations for a precise measurement at HERA.
\end{itemize}
The preliminary high x data from H1 is shown in figure \ref{highxvsq2} where the 
reduced cross section
$\sigma \approx F_2$ up to electroweak interference effects is shown versus $Q^2$
for different bins in x compared to BCDMS and SLAC data.
 Also shown is a QCD fit
 which used only low  $Q^2$ data.
The H1 measurements  at an average $<Q^2 > = 10000 \  GeV^2$ have an uncertainty of
 about 14\% limited mostly by the uncertainties in the energy calibration.
We therefore have so far directly checked the $Q^2$ evolution over 2 orders of
 magnitude to that precision. Further improvements will require better calibration
in addition to higher statistics. This is indeed possible at HERA. The energy 
calibration can be directly obtained from the data by using the redundancy in the kinematic
 reconstruction. At HERA we measure angle $\Theta _e $ and energy $E_e$
of the scattered electron as well as angle $\Theta _H$ and energy $E_H$ of the
hadronic system (the quark).  
\begin{figure}[htbt] \begin{center}
\epsfig{clip=,width=9.cm,file=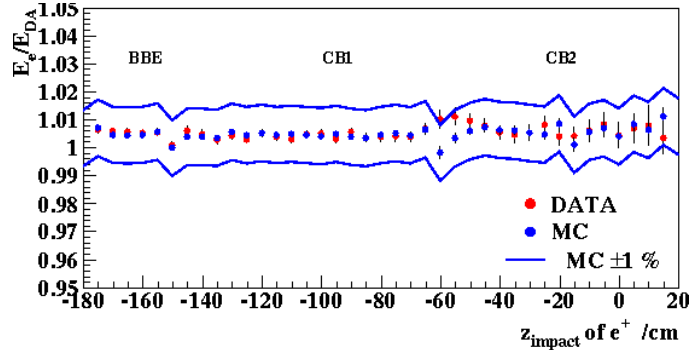}
 \caption{\it
Comparison of the H1 calibration of the electron energy in the liquid argon calorimeter
 compared to the energy form the  double angle method vs. the calorimeter position along
 the beam axis.
    \label{calibration} }
\end{center} \end{figure}
The energy  of the electron can therefore also be
calculated from the two angles alone $E_{DA} = f(\Theta _e, \Theta _H)$. The 
energy scale of the electromagnetic calorimeter vs. polar and azimuthal angle was
therefore  determined by comparison to the double angle energy. This comparison is shown
in figure \ref{calibration}.
 This procedure determines  the energy scale for the scattered
electron  to about 1\% for $\Theta _e > 15^o $ and to about 3\% for
smaller angles where the 3\% are  limited  by statistics. 
\par
HERA will substantially increase its luminosity  in 2000. High statistics
 data will
then  allow
better calibration and therefore very good and reliable checks for deviations
of the standard model. If a deviation should be observed due to new physics
then we are pretty sure that such an  effect could be established beyond doubt in
 contrast to similar
measurements at hadron colliders because only quarks can contribute and 
because the energy calibration is provided by the data themselves.
\clearpage
\section{ Global fits to parton distributions}
The determination of a complete set of parton distributions from available data
is the domain of two groups of phenomenologists, the CTEQ group centered at Fermilab
and the MRS group at Durham. They provide libraries of particle density functions (PDF)
which can be easily used and which are regularily updated if new data become available.
\par
Their global analyses use fit methods very similar to the one described above for H1. They
 use however additional input in order to also separate the flavour content of 
valence and sea quarks and to better constrain the gluon distribution. Main additional
 input comes from the following sources:
\begin{itemize}
\item Neutrino measurements from the CCFR collaboration \cite{CCFR} which determine \\
-the shape and magnitude of the valence distribution $x(u_v + d_v)$ from the
 structure function $xF_3$ \\
- the magnitude of the strange sea  $s(x)/(\bar u(x)+ \bar d(x)) \approx 0.25$ from single
 charmproduction  $\bar \nu + \bar s \rightarrow \mu^+ \bar c$.
- $\alpha_s $ determination from scaling violations of the nonsinglet structure function.
\item Data to determine u(x)/d(x) mostly in the valence region \\ - The direct NMC 
measurement of u(x)/d(x)  \cite{NMCu/d} and the measurement of the
asymmetry in W-production by CDF which is the most sensitive measurement at
large x
\cite{CDFW}.
\item The determination of $\bar u / \bar d$ by Drell-Yan experiments
\item Experiments measuring direct hard photon production in pN scattering as a direct
 measurement of the gluon distribution at large x. 
\end{itemize}
No attempt is made here to go into details. Comprehensive summaries can be found in
 references \cite{MRS} and \cite{CTEQ}. The results which are summarised  here
 in order to illustrate
sucesses and problems of this approach are taken from the recent fit of the MRS group
using the PDF set MRS98 \cite{MRS}. This basic PDF set fixes the value of
 $\alpha_s (m_Z)=.118$
to the world average.

 Global fits have  the problem that 
 data sets are  incompatible with each other and that practically all
 data sets are systematically limited with highly correlated errors. There is
no chance to include these errors - even if they would be fully available - into the
 fits.
This may lead to systematic biases of the results and it  makes it
 practically impossible to evaluate reliably the uncertainties
 of the parton densities.
 \subsection{ DIS data in the global fits}
\begin{figure}[htbt] \begin{center}
\epsfig{clip=,width=6.cm,file=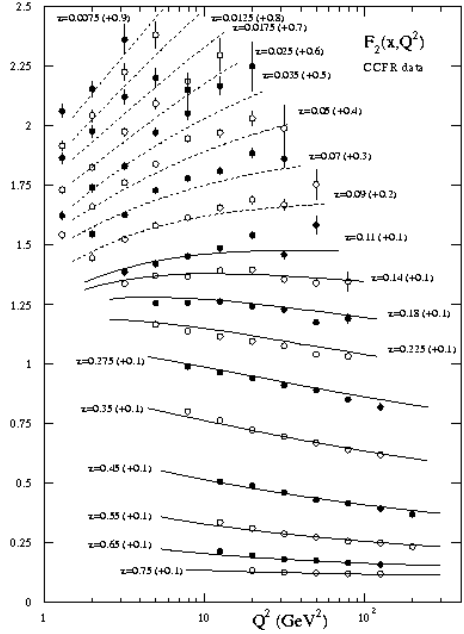 }
 \caption{\it
Measurement of $F_2^{\nu N} (x,Q^2) $ by the CCFR collaboration compared to MRS98 prediction.
Only measurements with $x > 0.1$ have been used in the fit.
    \label{CCFRf2} }
\end{center} \end{figure}
There is overall reasonable agreement. Some data sets give rather large contributions
to $\chi ^2$ / dof like e.g. the BCDMS ep and ed and the ZEUS structure
functions but the fits disregard the systematic errors.
Both MRS and CTEQ have decided to use single charm production from the CCFR neutrino
experiment to fix the strange sea. This however has a problem as illustrated in figure
\ref{CCFRf2}. The measured structure function $F_2^{\nu N}$ in the sea region ( x $< $
 0.1)
is 10 \% to 20 \% higher than expected whereas the  NMC muon data is well described.
Obviously one has to make a choice and MRS has decided NOT to use the low x CCFR data
 in the fit. There is however no explanation for this discrepancy.
\subsection{Drell-Yan processes and the $\bar u /\bar d$ ratio } 
\begin{figure}[htbt] \begin{center}
\epsfig{clip=,width=8.cm,file=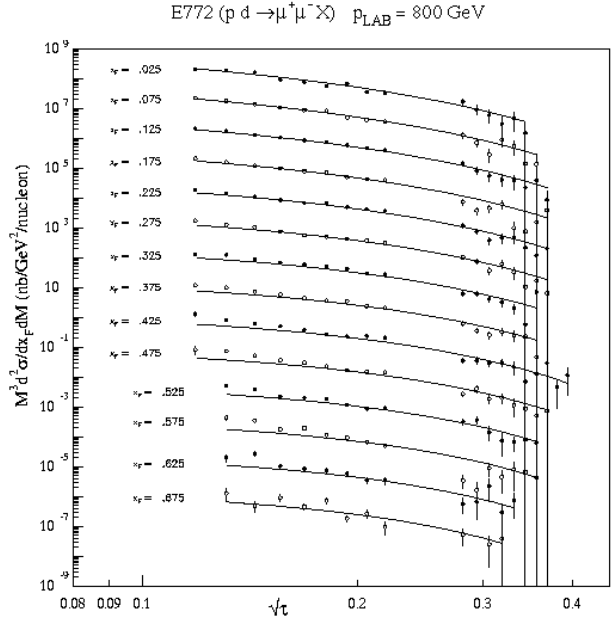}
 \caption{ \label{drellyan}     
\it Drell-Yan cross section of experiment E772 for the process 
 $pd \rightarrow \mu^+ \mu^- X$ \protect\cite{E772p} compared to the NLO predictions based on the
 MRS98 parametrisation }
\end{center} \end{figure}
\begin{figure}[htbt] \begin{center}
\epsfig{clip=,width=5.cm,file=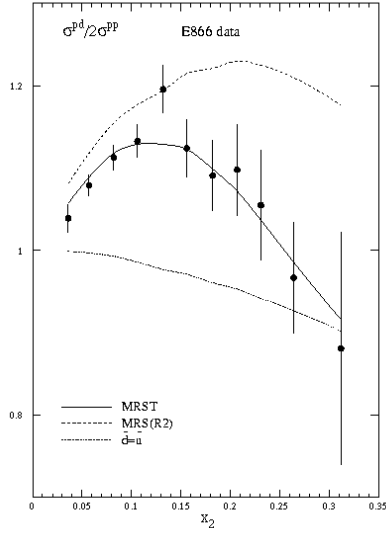 }
 \caption{\it The cross section ratio   $ \sigma (pd \rightarrow \mu^+ \mu_- X)/
2\star \sigma (pp \rightarrow \mu^+ \mu_- X)$ as measured by experiment E866
\protect\cite{E866}
compared to predictions based on $\bar u / \bar d =1.$ and on the MRS98 PDF with
$\bar d > \bar u $  \label{ubar/dbar} }
\end{center} \end{figure}
Lepton pair production in hadron-hadron scattering outside the $\Psi \ and \ Y $ mass
ranges as well as the production of the vector bosons at the $\bar p p $ collider are
rather well described by the Drell-Yan process. Figure \ref{drellyan} shows the 
differential cross section for the reaction $pd \rightarrow \mu^+ \mu^- X$ as measured
 by
 the E772 experiment at Fermilab at 800 GeV laboratory energy \cite{E772p}
 compared
 to NLO pQCD
predictions based on the MRS98 parametrisation.

 The overall agreement is quite good, 
at a level of about 10\%, except
in the region of large $x_F$ and small $\sqrt \tau $ where systematic deviations of up
 to a factor 2 are observed. This region corresponds to scattering processes where the
 first parton carries a momentum $x_1 \approx 0.4 \div 0.6 $ whereas the second parton is
 at small $x_2 \approx .02$. Both parton densities should therefore be well known such
 that the
 origin of the discrepancy is either an experimental problem or a problem 
in the theoretical
calculation.

 If one disregards this kind of problems then Drell-Yan data can be used to 
measure the flavour content of the sea quarks especially the $\bar u/\bar d $ ratio.
We know for quite some time that there are more down quarks in the sea than up quarks
because the Gottfried sum rule as measured by NMC \cite{NMCu/d} 
$$ \int_0^1 (F_2^{\mu p} - F_2^{\mu n})dx/x = 1/3 - 2/3 \int_0^1 (\bar d -\bar u)dx =
.235 \pm .026 $$ differs significantly from 1/3 and because the Drell-Yan experiment
NA51 measured $\bar d / \bar u > 1 \ at \ x \approx .5 $ \cite{NA51}. A new dedicated
experiment (E866) at Fermilab has made  very precise measurements in '97 to clarify this
question \cite{E866}. E866 measured the cross section ratio $ \sigma (pd \rightarrow
 \mu^+ \mu^- X)/
2\star \sigma (pp \rightarrow \mu^+ \mu^- X)$ directly which depends on $\bar d / \bar u$. 
This cross section ratio is shown in figure \ref{ubar/dbar} versus x compared
to predictions
for $\bar d/ \bar u =1 $. The  MRS98 fit uses this data to fix the flavour
ratio of the sea quarks.

The measured difference between down and up quarks in the sea is well compatible with
models where the nucleon has a pion cloud. \par
The ratio of up and down quark densities at large x e.g. in the valence region can be
derived from $F_2^{\mu n}/F_2^{\mu d}$ as measured by NMC \cite{NMCu/d}.  A very 
sensitive new 
measurement at high scale has been provided by CDF \cite{CDFW} which measured the 
asymmetry of the decay leptons from $W^{\pm}$ bosons at the $\bar p p $
collider. This
measurement  has
 been selected by MRS  to determine u/d in the valence region together of course 
with the BCDMS structure function measurements on proton and deuterium. 

\subsection{Measurement of the gluon distribution}

The gluon distribution is notoriously difficult to measure over the whole x range.
In the small x region ($x \le .1 $) the gluon distribution is reasonably well measured from
 $d F_2 /d ln Q^2$ by the HERA experiments and NMC as described above. This does not help
at larger x because there the scaling violations due to  quarks radiating gluons  dominate
 over the gluon splitting contribution. A very strong constraint for the gluon distribution at medium 
and large x
 is however given by the energy momentum sum rule.  \par
\begin{figure}[htbt] \begin{center}
\epsfig{clip=,width=6.cm,file=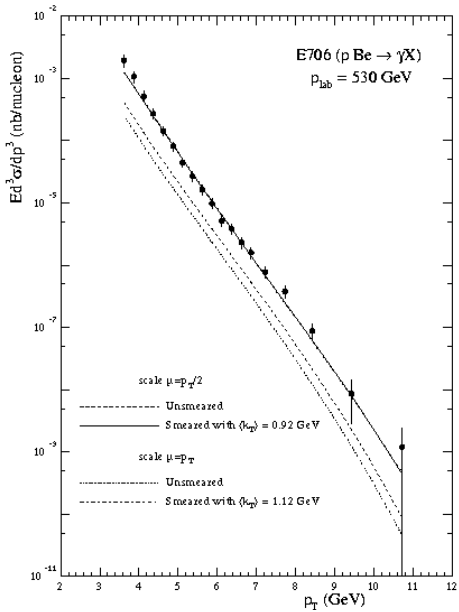}
 \caption{\it
The single photon cross section as measured by experiment E701 in pBe interactions
 at 530 GeV. The different curves give NLO predictions based on MRS98 using two scales
and adding an intrinsic $k_T=1.2 \ GeV$.
  \label{E706.1} }
\end{center} \end{figure}

A direct measurement of the gluon distribution at medium and large x is very
 desirable. The best suited process
is direct hard photon production in pN scattering. This process in leading order is
 absolutely dominated by quark-gluon scattering, $q \bar q $ processes give a relatively
small and well known contribution. Very precise new measurements for this process over
a large kinematic range have been provided by experiment E706 \cite{E706} as
shown in figure \ref{E706.1}.
 The invariant cross section for the process $p Be \rightarrow \gamma X $
at 530 GeV and 800 GeV 
has been measured for transverse momenta of the photon between  3.5 and 11 GeV/c which
 covers a range of the gluon fractional
 momentum $ .3 < x_{\gamma} < .5 $. The NLO pQCD prediction based on the 'canonical' gluon
distribution is unable to describe this data, it is about a factor 2 too low.
 The pQCD predictions have  a rather large scale  dependence; what is  however
 worst is the
 fact that single photon production requires a large 'intrinsic' $k_T \approx$ 1 GeV/c  of
 the incoming gluon
as e.g. shown by the imbalance of photon and jet momenta of $\gamma - jet$ events.
 Adding a value of $k_T = 1 \ GeV/c$ changes the cross section by about a factor 2
independent of $p_T$ 
\cite{E706}. Both
observations show that the theoretical understanding of this process is not
sufficient and as long as this does not change this process cannot be used to
rigoriusly
 constrain 
the gluon distribution at large x. This is rather unfortunate because these
 measurements can in
principle decide between conventional and 'unconventional'shapes of the
  gluon distribution at
 large x  which have been 
proposed  to  explain the
 excess of dijet events by CDF  \cite{CTEQcdf} .
\subsection{How well do we know the parton distributions?}
\begin{figure}[htbt] \begin{center}
\epsfig{clip=,width=7.cm,file=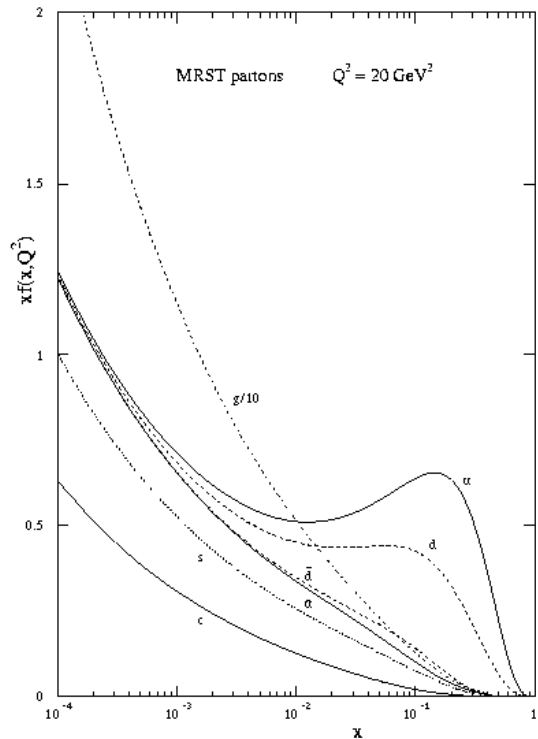}
 \caption{\it
Parton momentum distributions for a scale $Q^2=20 GeV^2$ for the MRS98 fit 
\protect\cite{MRS}.
  \label{partons20} } 
\end{center} \end{figure}

The parton momentum distributions from the MRS98 fit are shown in figure \ref{partons20}
for a scale $Q^2 = 20 \ GeV^2$.
\begin{figure}[htbt] \begin{center}
\epsfig{clip=,width=7.cm,file=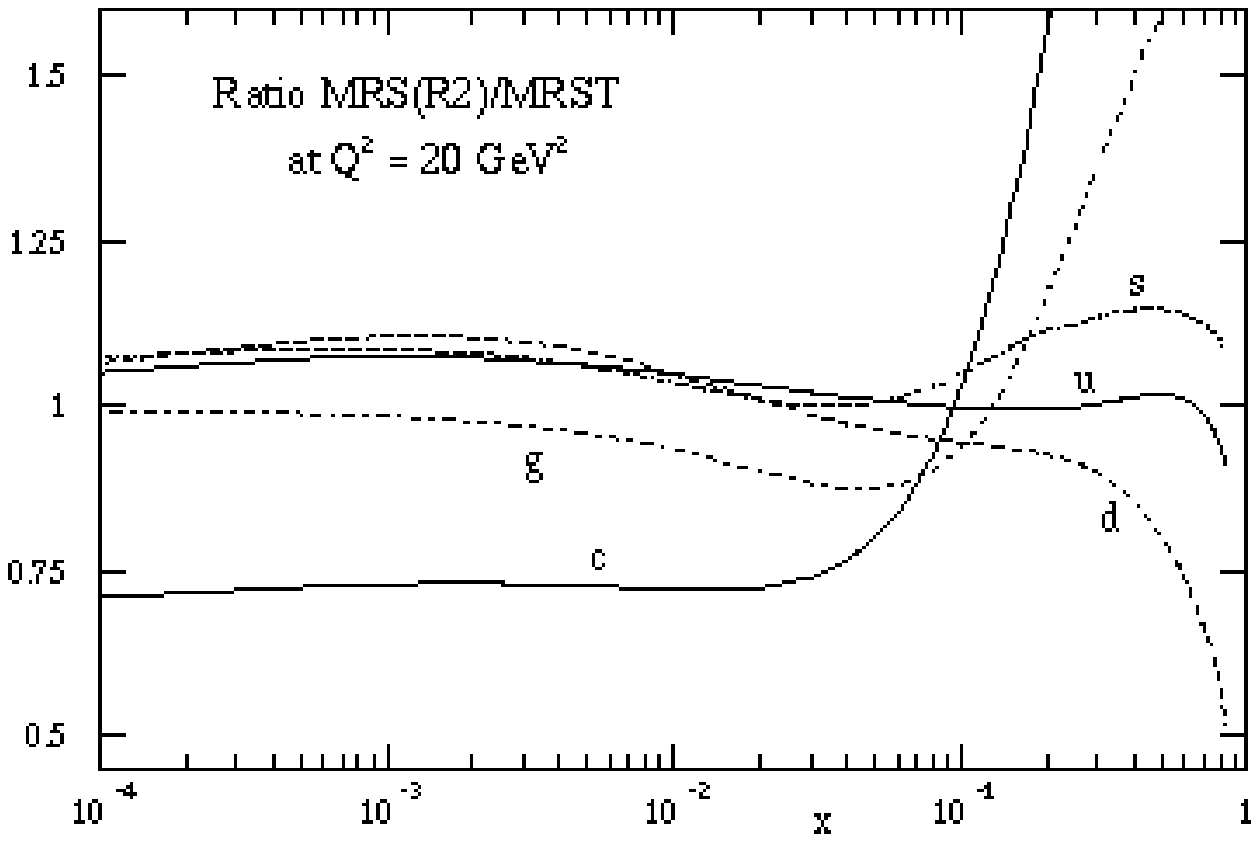}
 \caption{\it
Relative changes of parton densities vs. x for MRS fits from '97 and '98.
  \label{partonchange} } 
\end{center} \end{figure}
 These can be used to predict hard scattering processes. The question
 of course is: what are the errors for such a prediction. This information is not really 
available. The procedure to estimate this error by comparing different parton density
 parametrisations which can be found rather often is completely inadequate. 
 It happens too often that the addition of new data changes the PDF data sets far 
outside the error estimates given before. Examples of recent changes are given in figure 
\ref{partonchange} taken from the last MRS global fit paper \cite{MRS}.

 It shows the
relative changes for valence quarks and gluons for the MRS global fits of '97
and '98.
 Most surprising is the change of
 the down quark density at large x by up to 25 \%. This is the result of
 including the CDF $W^{\pm}$ asymmetry measurements in the fit.
 \par
The knowledge of parton distributions is experimentally limited by the
 fact that practically
all data sets are dominated by systematic errors where calibration errors are dominant. 
Different data sets also show inconsistencies, such that we have to make a choice which ones to use.
New data alone will not be sufficient to make progress. We need a better treatment of systematic
 errors in the fits where possible and we have to be more selective in choosing the data sets 
entering the fits. Those parts of data sets which have
large correlated systematic errors should be disregarded in future. \par
On the theoretical side more studies are needed to see if the parametrisations have enough
flexibility. NNLO calculations for DIS would be welcome in order to reduce the scale dependences.
Most of all we need however a satisfactory theoretical description of single photon
production to nail down the gluon distribution at large x. \par
Our knowledge of parton densities to my personal judgement can be roughly summarised as follows
 \begin{itemize}
\item the cross sections  for qq and $q\bar q$ processes
can be predicted with an error
$ \Delta \sigma /\sigma \le 10\%  \ for \ 10^{-4} < x < .3$ . The error increases to about 20 \%
 for larger values of x.
\item for gluon-gluon scattering we have $\Delta \sigma /\sigma \le 20\% \ for \ x < .1$.
 The error increases to 30\% up to x=.3 and is about 60\% for $.3 < x < .4$. These error estimates
are more conservative than what is normally given by the authors of PDF's.
\end{itemize}
Experimental improvements can be expected from HERA in future for quarks and
 gluon
 densities
 at small x with better statistics and improved systematic understanding of
 the detectors.
\clearpage
\subsection{ Determination of $\alpha _s$ from DIS.}
The uncertainty of $\alpha_s $ gives a non negligible contribution to our
knowlege of parton densities at high scales. 
The strong coupling constant  $\alpha _s$ can be well determined from the
observed scaling violations in DIS because the pQCD prediction
 does not suffer from large theoretical 
uncertainties and because experimental errors are also small. Since several
 years the world average of $\alpha _s$ 
from DIS experiments was always quoted to be low compared to LEP measurements.
 It should be noted that
this low value of  $\alpha _s$ was enforced by only two experiments: the BCDMS
 muon experiment which
 published a value  $\alpha _s (m_Z) = .113 \pm .003 _{exp} \pm .004_{th}. $ 
\cite{BCDMSfit} and by the CCFR neutrino
experiment which in '93 published a value
   $\alpha _s (m_Z) = .111 \pm .002 \pm .003_{syst}$ \cite{CCFRfit}
The CCFR collaboration has recently published a reanalysis of the same data \cite{CCFR}.
After significant improvements of the calibration for hadrons and muons the
 old published value
is changed to  $\alpha _s (m_Z) = .119 \pm .002_{exp} \pm .004_{th.}$.
 This leaves BCDMS as the
sole data set which requires a small value. Its worth while to have a closer
look to this data.
\begin{figure}[htbt] \begin{center}
\epsfig{clip=,width=6.cm,file=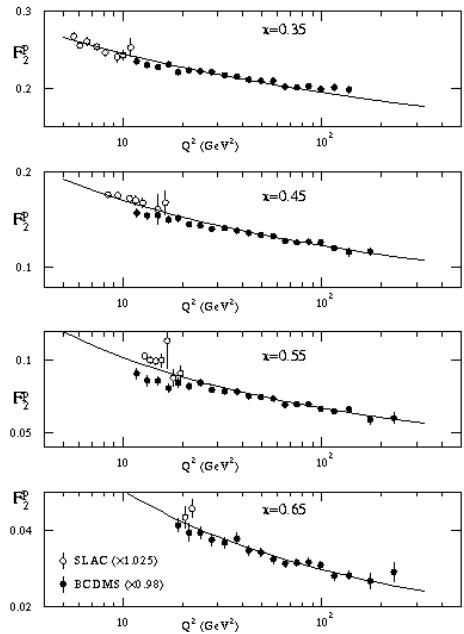}
 \caption{\it
The measurements of $F_2^{\mu d}$ at large x from BCDMS compared to
 neighbouring data points
 from SLAC and the MRS98 fit with  $\alpha _s = .118$.
  \label{BCDMSlargex} } 
\end{center} \end{figure}
The large x data for $F_2^{ed}$ from BCDMS is shown in figure
\ref{BCDMSlargex}
 together with SLAC data
 and a global QCD fit with $\alpha _s = .118$.
 For each x bin the data points at low $Q^2$ show large
 and very significant deviations from the fit curve but also from the SLAC
 data points leading 
to  a very poor $\chi ^2 / dof$ =248/174.
The data points with systematic deviations are  at low y where every DIS
 experiment
 has the most serious systematic
 problems due to energy calibration. This is also clearly stated in the BCDMS
 paper \cite{BCDMSed}
which says: ..'' the data agree with SLAC within the large correlated
 systematic errors..'. 
A fit which uses these statistically very precise data points disregarding the systematic
 errors will therefore severely bias the result and necessarily lead to wrong results.\par
My conclusion is that i) DIS data is in agreement with a world 
average of $\alpha _s \approx .118$
and ii) people which make global fits for PDF determination should care more about systematics,
best in collaboration with experimentalists.

\end{document}